\documentclass[prl,twocolumn,superscriptaddress,showpacs]{revtex4-2}
\usepackage{lineno}
\usepackage{amsmath}
\usepackage{units,xspace}
\usepackage{graphicx}
\usepackage{mathtools}
\usepackage{cuted}
\usepackage{color}
\usepackage{url,hyperref}
\usepackage{multirow}
\usepackage[utf8]{inputenc}
\usepackage{xcolor}
\newcommand{\ifb}{\mbox{fb$^{-1}$}\xspace}

\newcommand{\tev}{\mbox{TeV}\xspace}

\newcommand{\gd}{\ensuremath{g_{\textrm D}}\xspace}

\newcommand{\half}{\nicefrac{1}{2}\xspace}
\newcommand{\MAD}{\textsc{MadGraph5}\xspace}
\newcommand{\geant}{\textsc{Geant4}\xspace}

\DeclareGraphicsExtensions{.pdf,.png,.jpg}
\graphicspath{{./}{./fig/}}

\begin{document}


\sloppy



\newpage

\title{First search for dyons with the full MoEDAL trapping detector \\
in $13~\tev$ $pp$ collisions}

\author{B.~Acharya}
\altaffiliation[Also at ]{Int. Centre for Theoretical Physics, Trieste, Italy}   
\affiliation{Theoretical Particle Physics \& Cosmology Group, Physics Dept., King's College London, UK}

\author{J.~Alexandre}
\affiliation{Theoretical Particle Physics \& Cosmology Group, Physics Dept., King's College London, UK}


\author{P.~Benes}
\affiliation{IEAP, Czech Technical University in Prague, Czech~Republic}

\author{B.~Bergmann}
\affiliation{IEAP, Czech Technical University in Prague, Czech~Republic}

\author{J.~Bernab\'{e}u}
\affiliation{IFIC, Universitat de Val\`{e}ncia - CSIC, Valencia, Spain}

\author{A.~Bevan}
\affiliation{School of Physics and Astronomy, Queen Mary University of London, UK}

\author{H.~Branzas}
\affiliation{Institute of Space Science, Bucharest - M\u{a}gurele, Romania}

\author{P. Burian}
\affiliation{IEAP, Czech Technical University in Prague, Czech~Republic}

\author{M.~Campbell}
\affiliation{Experimental Physics Department, CERN, Geneva, Switzerland}

\author{S.~Cecchini}
\affiliation{INFN, Section of Bologna, Bologna, Italy}

\author{Y.~M.~Cho}
\affiliation{Center for Quantum Spacetime, Sogang University, Seoul, Korea} 

\author{M.~de~Montigny}
\affiliation{Physics Department, University of Alberta, Edmonton, Alberta, Canada}

\author{A.~De~Roeck}
\affiliation{Experimental Physics Department, CERN, Geneva, Switzerland}

\author{J.~R.~Ellis}
\altaffiliation[Also at ]{National Institute of Chemical Physics \& Biophysics, Tallinn, Estonia} 
\affiliation{Theoretical Particle Physics \& Cosmology Group, Physics Dept., King's College London, UK}
\affiliation{Theoretical Physics Department, CERN, Geneva, Switzerland}

\author{M.~El~Sawy}
\altaffiliation[Also at ]{Dept. of Physics, Faculty of Science, Beni-Suef University, Egypt} 
\affiliation{Experimental Physics Department, CERN, Geneva, Switzerland}

\author{M.~Fairbairn}
\affiliation{Theoretical Particle Physics \& Cosmology Group, Physics Dept., King's College London, UK}

\author{D.~Felea}
\affiliation{Institute of Space Science, Bucharest - M\u{a}gurele, Romania}

\author{M.~Frank}
\affiliation{Department of Physics, Concordia University, Montr\'{e}al, Qu\'{e}bec,  Canada}

\author{J.~Hays}
\affiliation{School of Physics and Astronomy, Queen Mary University of London, UK}

\author{A.~M.~Hirt}
\affiliation{Department of Earth Sciences, Swiss Federal Institute of Technology, Zurich, Switzerland}

\author{J.~Janecek}
\affiliation{IEAP, Czech Technical University in Prague, Czech~Republic}

\author{M.~Kalliokoski}
\affiliation{Physics Department, University of Helsinki, Helsinki, Finland}


\author{A.~Korzenev}
\affiliation{D\'epartement de Physique Nucl\'eaire et Corpusculaire, Universit\'e de Gen\`eve, Geneva, Switzerland}

\author{D.~H.~Lacarr\`ere}
\affiliation{Experimental Physics Department, CERN, Geneva, Switzerland}


\author{C.~Leroy}
\affiliation{D\'{e}partement de Physique, Universit\'{e} de Montr\'{e}al, Qu\'{e}bec, Canada}

\author{G.~Levi} 
\affiliation{INFN, Section of Bologna \& Department of Physics \& Astronomy, University of Bologna, Italy}

\author{A.~Lionti}
\affiliation{D\'epartement de Physique Nucl\'eaire et Corpusculaire, Universit\'e de Gen\`eve, Geneva, Switzerland}

\author{J.~Mamuzic}
\affiliation{IFIC, Universitat de Val\`{e}ncia - CSIC, Valencia, Spain}

\author{A. Maulik}
\affiliation{INFN, Section of Bologna, Bologna, Italy}
\affiliation{Physics Department, University of Alberta, Edmonton, Alberta, Canada}

\author{A.~Margiotta}
\affiliation{INFN, Section of Bologna \& Department of Physics \& Astronomy, University of Bologna, Italy}

\author{N.~Mauri}
\affiliation{INFN, Section of Bologna, Bologna, Italy}

\author{N.~E.~Mavromatos}
\affiliation{Theoretical Particle Physics \& Cosmology Group, Physics Dept., King's College London, UK}

\author{P.~Mermod}
\altaffiliation[Now deceased ]{} 
\affiliation{D\'epartement de Physique Nucl\'eaire et Corpusculaire, Universit\'e de Gen\`eve, Geneva, Switzerland}

\author{M.~Mieskolainen}
\affiliation{Physics Department, University of Helsinki, Helsinki, Finland}

\author{L.~Millward}
\affiliation{School of Physics and Astronomy, Queen Mary University of London, UK}

\author{V.~A.~Mitsou}
\affiliation{IFIC, Universitat de Val\`{e}ncia - CSIC, Valencia, Spain}

\author{R.~Orava}
\affiliation{Physics Department, University of Helsinki, Helsinki, Finland}

\author{I.~Ostrovskiy}
\affiliation{Department of Physics and Astronomy, University of Alabama, Tuscaloosa, Alabama, USA}

\author{P.-P. Ouimet}
\altaffiliation[Also at ]{Physics Department, University of Regina, Regina, Saskatchewan, Canada}
\affiliation{Physics Department, University of Alberta, Edmonton, Alberta, Canada}
  
\author{J.~Papavassiliou}
\affiliation{IFIC, Universitat de Val\`{e}ncia - CSIC, Valencia, Spain}

\author{B.~Parker}
\affiliation{Institute for Research in Schools, Canterbury, UK}

\author{L.~Patrizii}
\affiliation{INFN, Section of Bologna, Bologna, Italy}

\author{G.~E.~P\u{a}v\u{a}la\c{s}}
\affiliation{Institute of Space Science, Bucharest - M\u{a}gurele, Romania}

\author{J.~L.~Pinfold}
\email[Corresponding author: ]{jpinfold@ualberta.ca}
\affiliation{Physics Department, University of Alberta, Edmonton, Alberta, Canada}

\author{L.~A.~Popa}
\affiliation{Institute of Space Science, Bucharest - M\u{a}gurele, Romania}

\author{V.~Popa}
\affiliation{Institute of Space Science, Bucharest - M\u{a}gurele, Romania}

\author{M.~Pozzato}
\affiliation{INFN, Section of Bologna, Bologna, Italy}

\author{S.~Pospisil}
\affiliation{IEAP, Czech Technical University in Prague, Czech~Republic}

\author{A.~Rajantie}
\affiliation{Department of Physics, Imperial College London, UK}

\author{R.~Ruiz~de~Austri}
\affiliation{IFIC, Universitat de Val\`{e}ncia - CSIC, Valencia, Spain}

\author{Z.~Sahnoun}
\altaffiliation[Also at ]{Centre for Astronomy, Astrophysics and Geophysics, Algiers, Algeria}
\affiliation{INFN, Section of Bologna, Bologna, Italy}

\author{M.~Sakellariadou}
\affiliation{Theoretical Particle Physics \& Cosmology Group, Physics Dept., King's College London, UK}

\author{A.~Santra}
\affiliation{IFIC, Universitat de Val\`{e}ncia - CSIC, Valencia, Spain}

\author{S.~Sarkar}
\affiliation{Theoretical Particle Physics \& Cosmology Group, Physics Dept., King's College London, UK}

\author{G.~Semenoff}
\affiliation{Department of Physics, University of British Columbia, Vancouver, British Columbia, Canada}

\author{A.~Shaa}
\affiliation{Physics Department, University of Alberta, Edmonton, Alberta, Canada}

\author{G.~Sirri}
\affiliation{INFN, Section of Bologna, Bologna, Italy}

\author{K.~Sliwa}
\affiliation{Department of Physics and Astronomy, Tufts University, Medford, Massachusetts, USA}

\author{R.~Soluk}
\affiliation{Physics Department, University of Alberta, Edmonton, Alberta, Canada}

\author{M.~Spurio}
\affiliation{INFN, Section of Bologna \& Department of Physics \& Astronomy, University of Bologna, Italy}

\author{M.~Staelens}
\affiliation{Physics Department, University of Alberta, Edmonton, Alberta, Canada}

\author{M.~Suk}
\affiliation{IEAP, Czech Technical University in Prague, Czech~Republic}

\author{M.~Tenti}
\affiliation{INFN, CNAF, Bologna, Italy}

\author{V.~Togo}
\affiliation{INFN, Section of Bologna, Bologna, Italy}

\author{J.~A.~Tuszy\'{n}ski}
\affiliation{Physics Department, University of Alberta, Edmonton, Alberta, Canada}

\author{A.~Upreti}
\affiliation{Department of Physics and Astronomy, University of Alabama, Tuscaloosa, Alabama, USA}

\author{V.~Vento}
\affiliation{IFIC, Universitat de Val\`{e}ncia - CSIC, Valencia, Spain}

\author{O.~Vives}
\affiliation{IFIC, Universitat de Val\`{e}ncia - CSIC, Valencia, Spain}

\author{A.~Wall}
\affiliation{Department of Physics and Astronomy, University of Alabama, Tuscaloosa, Alabama, USA}

\collaboration{THE MoEDAL COLLABORATION}
\noaffiliation

\date{\today}

\begin{abstract}
The MoEDAL trapping detector, consists of approximately 800~kg of aluminium volumes. It was exposed during Run-2 of the LHC program  to 6.46~\ifb of 13~\tev proton-proton collisions at the LHCb interaction point.  Evidence for  dyons (particles with electric and magnetic charge)  captured in the trapping detector was sought by passing the aluminium volumes comprising the detector through a SQUID magnetometer. The presence of a trapped dyon would be signalled by a  persistent current induced in the SQUID magnetometer.  On the basis of a Drell-Yan production model, we exclude dyons with a magnetic charge ranging up  to 5 Dirac charges (5$\gd$)  and an electric charge up to 200 times the fundamental electric charge for mass limits in the range 750 -- 1910~GeV, and also monopoles with magnetic charge up to and including 5$\gd$ with mass limits in the range 870 -- 2040~GeV. \\ 
\\
\emph{This paper is dedicated to the memory of Philippe Mermod, a founding and leading member of the MoEDAL experiment.}
\end{abstract}


\maketitle


The  search for the magnetic monopole has been a key concern of  fundamental physics since  Dirac in 1931 \cite{Dirac:1931kp}  demonstrated its existence was consistent with quantum mechanics provided the  quantization condition (in  SI units) $g/e = n(c/2\alpha_{em}$) is satisfied, where $g$ is the magnetic charge, $e$ is a unit electric charge,  $c$ is the speed of light, $\alpha_{em}$ is the fine structure constant, and $n$ is an integer.  When $n = 1$ then  $g = \gd$,  one  Dirac charge.
There is a long history of direct  searches for magnetic monopoles at accelerators~\cite{PDG18}, most recently at the LHC~\cite{MoEDAL:2016,Acharya:2017,Acharya:2018,Acharya:2019,Aad12,Aad16,Aad19}, and there have also been extensive searches for monopole relics from the early Universe in cosmic rays and in materials~\cite{Burdin15,Patrizii15,Mavromatos:2020gwk}.

The existence of  the dyon, a particle  with both magnetic and electric charge, was first proposed by Julian Schwinger in 1969 \cite{Schwinger1969}. Schwinger derived the following charge quantization condition by considering the interaction of two dyons:
\begin{equation}
e_{1}g_{2} - e_{2}g_{1} = \frac{n}{2}\hbar c
\end{equation}
where, $e_{1}$,  $e_{2}$  and $g_{1}$, $g_{2}$ are the electric and magnetic charges of the two dyons, respectively. This quantization condition does not, by itself, fix the electric charge of the dyon, and provides no $a$  $priori$ limitation on the size of the electric charge of the dyon.
 However, the issue of the charge of the quantum dyon has been studied carefully by semi-classical reasoning \cite{goldstone76}, and it has been concluded that in CP-conserving theories the dyon charge is quantized as an integer multiple of the fundamental charge, $q = ne$.  
 
When the theory admits CP non-conservation this is no longer the case. The  topologically nontrivial vacuum structure of  non-Abelian gauge theories is characterized by the vacuum angle $\theta$, or ``theta term'', which can be added to the Lagrangian for Yang-Mills theory without spoiling renormalizability. Witten~\cite{witten79} considered CP violation induced by a vacuum angle in the context of  the Georgi-Glashow model that gives rise to the  non-Abelian monopole of 't Hooft and Polyakov, showing that dyons are magnetic monopoles with fractional electric charge. He derived the following  relation between the dyon's electric charge and $\theta$:
\begin{equation}
q = ne - \frac{e\theta}{2\pi} \, .
\end{equation}
Experiment has found that CP is only weakly violated. As the deviation of the monopole from integral charge is proportional to the strength of CP violation, one would therefore expect the dyon charge to have almost, but not quite,  an integer value. 
 \vspace*{-3mm}
\begin{figure}[htb]
\begin{center}
  \includegraphics[width=0.5\linewidth]{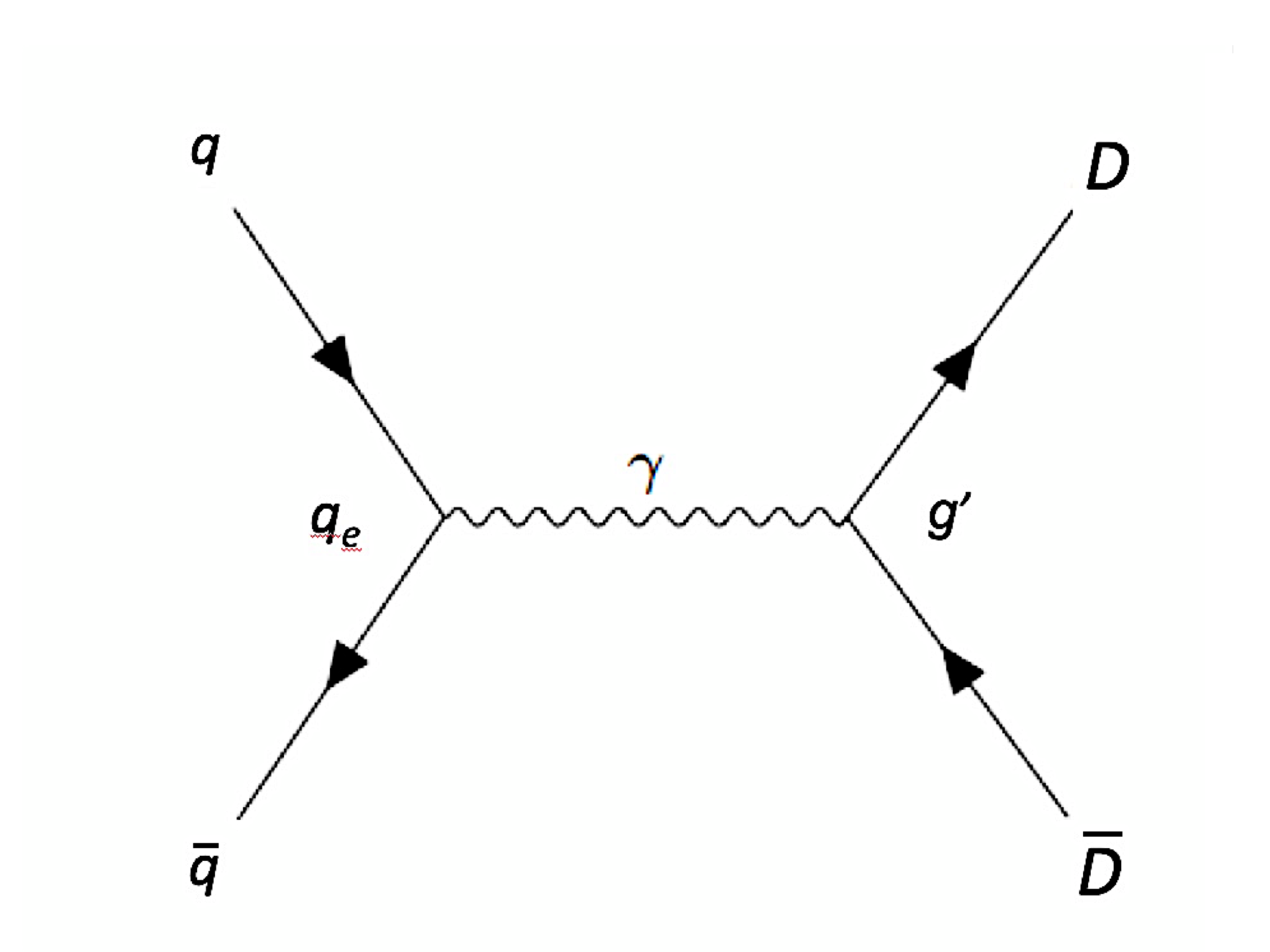}
 \vspace*{-5mm}
  \caption{Feynman-like diagram for dyon-pair direct production at leading order via the benchmark Drell-Yan mechanism. 
  The coupling $g'$ is given by $\sqrt{(g^{2} + q^{2})}$,  in the eikonal approximation that is valid for LHC energies.}
 \vspace*{-5mm}
\label{fig:diagrams}
\end{center}
\end{figure}

Since Schwinger's original work, it has been shown that 
dyons appear generically in theories with monopoles, specifically in many particle-physics theories including Grand Unified Theories (GUTs)   \cite{GUT-dyons}, Einstein-Yang-Mills theories \cite{YM-dyons}, Kaluza-Klein theory \cite{KK-dyons}, string theory \cite{String-dyons}, and M-theory \cite{Mtheory-dyons}.  Moreover, a number of theoretical scenarios have been proposed that contain Electroweak (EW) dyons and monopoles \cite{ymincho-2019,hung-2020,am-2019,ms-2018,yme-2016} that could be detected at the LHC or the High-Luminosity LHC (HL-LHC) \cite{HE-LHC}. We note also that the production of such EW dyons and monopoles during the EW phase transition in the early Universe would have major implications for cosmology~\cite{ymincho-2019}.



The electric charge of a dyon can, in principle,  be large. Other suggested examples of
highly-electrically-charged  (pseudo-) stable objects (HECOs) include: aggregates of $ud$- \cite{Holdom18} or $s$-quark matter \cite{Farhi84}, $ Q$-balls \cite{Coleman85,Kusenko98} and the remnants of microscopic black-holes \cite{Koch07}. Extensive accelerator searches  for HECOs have also been undertaken \cite{Aad11,Aad16,burdin:2015}.
Recently ATLAS has placed improved limits on HECOs
\cite{Aad19},  extending a previous excluded charge range from 20$ \le |z| \le$ 60  \cite{Aad16} to 60 $ \le |z| \le$ 100 \cite{Aad19}, where $z$ is the electric charge~\cite{Screening}.
However, we are unaware of any previous accelerator search for dyons.

The Drell-Yan  (DY) production mechanism shown in Fig.~\ref{fig:diagrams} is frequently employed in accelerator-based searches for monopoles~\cite{Aad12,Aad16, Aad19,MoEDAL:2016,Acharya:2017,Acharya:2018,Acharya:2019} and provides a simple benchmark model of monopole-pair production.  Here we use a similar DY production model also as a benchmark for dyon production. As in the previous MoEDAL monopole searches~\cite{Acharya:2018, Acharya:2019}, spins of  0, \half and~1 are considered  and models were generated in \MAD~\cite{Alwall:2014} using the Universal FeynRules Output described in  Ref.~\cite{Baines:2019}. We used tree-level  diagrams and the parton distribution functions \texttt{NNPDF23}~\cite{Ball:2013} for the DY  process.  The square of the magnetic  charge of the monopole,   $g^{2}$, is substituted in the basic DY cross-section  by $g^{2} + q^{2}$, where $q$ is the electric charge of the dyon defined above.   This scaling  is in accord with the  dual effective theory of Milton and Gamberg \cite{Milton:2000,Milton:2000x} and the theoretical approach developed for monopoles  in Ref.~\cite{Baines:2019} when extended to dyons.

There are two important differences between the signatures of the magnetic monopole and dyon at the LHC that are due to the electric charge of the dyon. First, a relativistic monopole, with  magnetic charge $n\gd \, (n = 1,2,3...)$ and  fractional velocity $\beta = v/c$, where $v$ is the monopole velocity,
behaves like an equivalent electric charge $(Ze)_{eq} =n\gd\beta$ where  $Z$  is the effective ``atomic number''. The energy loss of a fast monopole is thus very large with a different velocity dependence from that of an electrically-charged particle.  On the other hand, the ionization energy loss of a  dyon is  the sum of the energy loss due to its electric charge and magnetic  charge, with their different velocity dependences.

Secondly, the magnetic monopole follows a curved trajectory in  the $r - z$ plane of a solenoidal field, where $z$ is the direction of the field lines, and $r$ is the radial dimension, without bending in the transverse plane. This is opposite to  the behaviour of an electrically-charged particle in the same field. On the other hand, the trajectory of a dyon in a solenoidal field curves in both the $r-z$ plane and in the plane transverse to this plane. Thus, its trajectory is distinct from that of either an isolated electric or magnetic charge.

The response of MoEDAL to
the passage of a monopole or a dyon differs significantly from those of the general-purpose LHC experiments, ATLAS and CMS. The MoEDAL detector, deployed along with LHCb at LHC intersection point IP8,   employs two unconventional passive detection methodologies tuned to the discovery of highly-ionizing particles (HIPs).  The first, used in this analysis, utilizes a 800 kg trapping detector (MMT) comprised of 2400 aluminium (Al) bars to capture HIPs for further study. The second consists of an array of 186 Nuclear Track Detector stacks. MMT volumes are deployed just upstream and on each side of IP8 in three roughly equal masses each placed 1.0 m - 2.0 m from the IP.  After exposure the MMTs' Al bars are passed through a SQUID magnetometer at the  ETH Zurich Laboratory for Natural Magnetism in order to check for the presence of magnetic charge. Further information on  the MoEDAL  detector is given   in the supplemental material \cite{supplemental} which includes Refs    \cite{LHCb, Royal-Society}.

To date, only the ATLAS and MoEDAL experiments have reported limits on monopole production at the LHC \cite{Aad12,Aad16, Aad19,MoEDAL:2016,Acharya:2017,Acharya:2018,Acharya:2019}. MoEDAL's latest search results \cite{Acharya:2019} include the combined photon-fusion and DY monopole-pair production mechanisms, the former process for the first time at the LHC. Using 4.0~\ifb of data, cross-section upper limits as low as 11 fb were set, and mass limits in the range 1500 - 3750 GeV were set for magnetic charges up to 5$g_{D}$ for monopoles of spins 0, 1/2, and  1, the strongest to date at a collider experiment. These limits are based on a  {\it direct} search for magnetic charge, with an unambiguous signature.

The most recent ATLAS search \cite{Aad19} placed 95\% confidence level mass limits  on DY production of spin-0 and spin-1/2 monopoles,  with charge 1$\gd$,  of 1850 GeV  and 2370 GeV, respectively. The corresponding ATLAS limits for charge 2$g_{D}$ monopoles are 1725 GeV and 2125 GeV, respectively. For magnetic charge $g_{D} \le$ 2 these are currently the world's best limits based on the  ionizing nature of magnetic monopoles or dyons.

\if
Dyons also possess electric charge that can, in principle,  be large.
The existence of highly-charged  (pseudo-) stable HECOs has also been hypothesized. Examples of such phenomena are: aggregates of $ud$- \cite{Holdom18} or $s$-quark matter \cite{Farhi84}, $ Q$-balls \cite{Coleman85,Kusenko98} and the remnants of microscopic black-holes \cite{Koch07}. Extensive accelerator searches  for HECOs have also been undertaken \cite{Aad11,Aad16,burdin:2015}.
Recently ATLAS has placed improved limits on  stable objects with electric charge \cite{Aad19}  extending a previous excluded charge range from 20$ \le |z| \le$ 60  \cite{Aad16} to 60 $ \le |z| \le$ 100 \cite{Aad19}, where $z$ is the electric charge. 

In the search for HECOs one must consider the  possibility of screening due to vacuum pair production via the Schwinger mechanism in the case of high electric charges. As argued in \cite{englert}, screening can affect the results for electric charges higher than 500, due to the strong electric fields produced in such cases. 
\fi

A monopole is expected to be stopped when  its velocity falls to  $\beta \le$ 10$^{-3}$ and then bind, due to interaction between the
monopole and the nuclear magnetic moment \cite{binding,goebel:1984,bracci:1984,olaussen:1985}. The large magnetic moment  gives a predicted monopole-nucleus
binding energy (BE)  of 0.5 -  2.5 MeV \cite{binding}. These BEs
are comparable to the shell model splittings. Thus, it is
reasonable to assume  that, in any case, the strong magnetic
field in the monopole's vicinity will rearrange the nucleus allowing
 it to strongly bind to the nucleus. According to Ref. \cite{binding}
monopoles with this  BE will be bound indefinitely,
requiring  fields  in excess of around 5T for them to be  released. We note in this connection that the MMT  volumes are never subjected to such strong
magnetic fields.

The dyon is also expected to stop when  its velocity falls to  $\beta \le$ 10$^{-3}$.
However, the binding of the dyon
is complicated by its electric  charge. In our analysis, we assume conservatively that only dyons with negative electric charge  are bound, since in this case their Coulomb attraction to the positive charge of the nucleus
reinforces the interaction between its magnetic charge and the  large anomalous nuclear magnetic moment of the aluminium nucleus.
Although the trapping condition requires the dyon to be negatively electrically  charged, the assumption of DY production of dyon  - antidyon pairs implies indirect sensitivity to positively-charged dyons at the same level.

A magnetic charge captured  in a trapping volume bar is identified and  measured as a persistent current in the coil of the SQUID surrounding the transport axis of the MMTs' Al bars. 
The magnetic pole strength, expressed in units of the Dirac charge,  contained in a sample is calculated as $P=C\cdot\left[(I_2-I_1)-(I^{\text{tray}}_2-I^{\text{tray}}_1)\right]$, where $C$ is the calibration constant;  ($I_1$)  ($I_2$) are the currents measured before and after the sample has passed through the sensing coil; and, $I^{\text{tray}}_2$ and $I^{\text{tray}}_1$ are the corresponding contributions measured with an empty tray.
The empty tray contributions arise from small seemingly random fluctuations  of the SQUID measurement due largely to less than  perfect grounding of the SQUID magnetometer electronics. It should be noted that the small tray is constructed from G10, a non-metallic and non-magnetic fibreglass-epoxy composite  that cannot shield or enhance the magnetic signal. 

The magnetometer response is calibrated  using two independent methods, described in more detail in Ref.~\cite{DeRoeck}. 
The two methods agree within 10\%, which we take  as the calibration uncertainty in the pole strength. The magnetometer response is measured to be linear and charge-symmetric in a range corresponding to 0.3 - 300 $\gd$.  During Run-2  the plateau value of the calibration dipole sample was remeasured regularly  and found to be stable to within less than 1\%. In 2018, the SQUID was overhauled, the main improvement being better grounding throughout the SQUID magnetometer mechanics and electronics. This had the effect of substantially reducing  fluctuations in the recorded magnetometer values.


Each MMT sample was scanned at least twice. A sample containing a dyon would repeatedly and  consistently yield the same non-zero measurements corresponding to the magnetic charge of the dyon. When a dyon is not present values  consistent with zero would be recorded. If  the measured pole strength of a sample differed from zero by more than $0.4\gd$ in either of the two initial measurements, it was considered a candidate. In this way the probability of false negatives was significantly reduced. A total of 87 candidates   were thus identified in data taken in 2015, 2016 and 2017, corresponding to 4.0 fb$^{-1}$.  
Only  29  candidates were observed in the data 2018 data (Run-B), where 2.46 fb$^{-1}$ of luminosity was recorded.  The MMT volumes containing  dyon  candidates were  rescanned  several times. For each candidate it was found  that the majority of the pole strengths measured  were below the threshold of $0.4\gd$.


The maximum probability for missing a dyon in a single measurement was found to be 0.53\% for a charge of $\pm$1$\gd$. As two passes were made for each sample during Run~A (2015 - 2017)  we have the negligible  probability of missing the dyon twice of 0.0028\%. Also, in  Run-A data it was found that candidate events were associated with greater than average fluctuations in the SQUID signal. In this case the probability of missing a dyon candidate was determined, using the 87 candidate events in Run-A,  to be 0.2\%. These  probabilities  become smaller with increasing magnetic charge. 
A more detailed description of the estimation of these probabilities in given in the supplementary information \cite{supplemental}.
In order to make a conservative estimate, we did not use  the Run-B  data to assess the probability of missing a dyon.


We define the acceptance for the MMT detector to be the fraction of the number of events in which at least one dyon of the pair in an event was trapped in the MoEDAL trapping detector. The trapping condition is determined from the knowledge of the material traversed by the dyon \cite{Alves:2008,MoEDAL:2016} and the ionization energy loss of dyons when they go through matter~\cite{Ahlen:1978,Ahlen:1980,Ahlen:1982,Cecchini:2016}, implemented in a simulation based on \geant~\cite{Allison:2006}. 
For a given dyon mass and charge, the pair-production model determines the kinematics and the overall trapping acceptance obtained. The uncertainty in the acceptance is dominated by uncertainties in the material description~\cite{MoEDAL:2016,Acharya:2017,Acharya:2018}. This contribution is estimated by performing simulations with hypothetical material conservatively added and removed from the nominal geometry model.

There are three causes of acceptance loss. The first is due to the limited geometrical extent of the MMT detector and the spin dependence in the geometrical acceptance due to the different event kinematics. The  second loss of acceptance is due to heavier, slower, dyons  with smaller effective ionizing power  punching through the trapping detector. We recall that in the case of magnetic charge the energy loss per unit distance falls with velocity. The third cause of acceptance loss is due to  the dyon being absorbed in the material comprising the VELO detector, which  encompasses the interaction point, before it reaches the MMT trapping volumes.

The largest acceptance is for dyons with spin-1 and magnetic charge 2$\gd$  where, for mass up to $\sim$3 TeV and  electric charges up to $\sim$50$e$, the acceptance is  greater than or equal to 2.1\%. 
 The acceptance is  below 0.1\% over the whole mass range considered, for dyons 
 that carry a magnetic charge of 6$\gd$ or greater, for all values of electric charge. 
The maximum dyon electric and magnetic  charge to which this analysis is sensitive is $\sim$200$e$ and 5$\gd$, respectively.
 
 The material encountered by particles within the acceptance of the MoEDAL, before they reach the MoEDAL detector varies from 0.1 to 8 radiation lengths ($X_{0}$) with an average of roughly 1.4$X_{0}$. The dominant systematic uncertainty comes from the estimated amount of material in the \geant geometry description, yielding a relative uncertainty of $\sim10\%$ for 1$\gd$ dyons~\cite{MoEDAL:2016}. This uncertainty increases with the magnetic and electric charge, reaching a point (at 6$\gd$) where it is too large for the analysis to be meaningful for spin-0 and spin-1/2 dyons. But, limits can be placed for spin-1 dyons with magnetic charge 6$\gd$ and electric charge from 1 to ~50$e$.

\begin{figure*}[!ht]
  \begin{center}
    \includegraphics[width=0.495\linewidth]{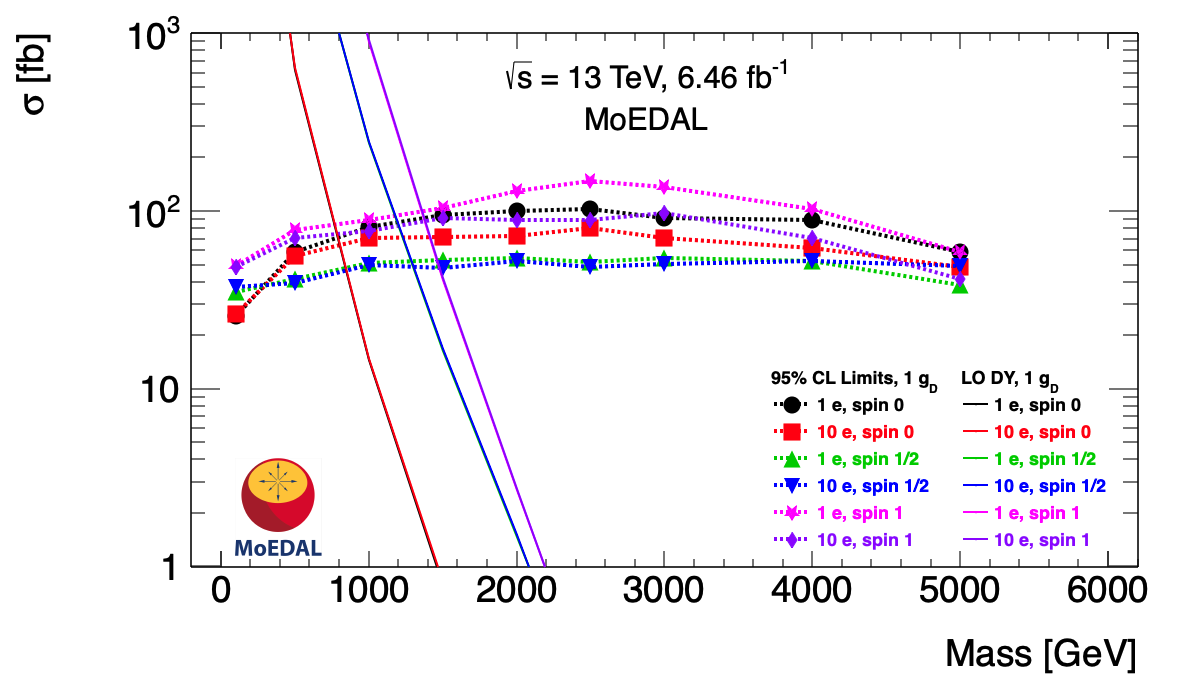}  
    \includegraphics[width=0.495\linewidth]{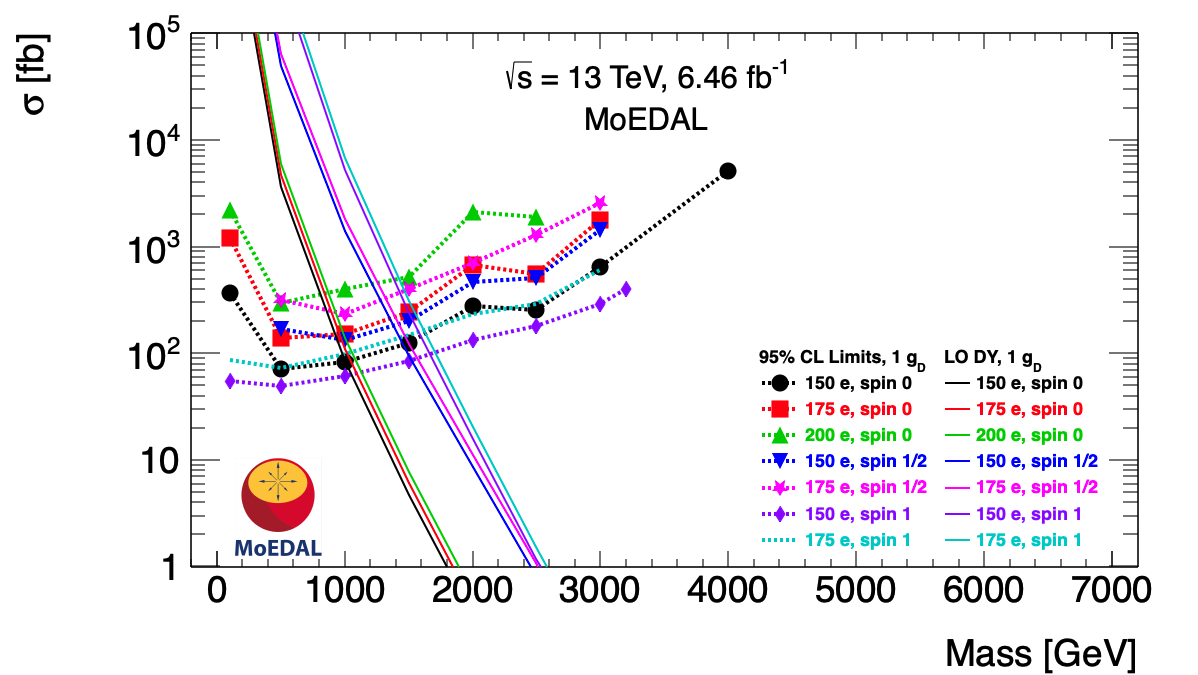}  
    \end{center}
    \vspace*{-7mm}
  \caption{\label{fig:limits1} Cross-section upper limits at 95\% CL for  DY spin-0, 1/2 and 1  dyon-pair production, with magnetic charge 1$g_{D}$ and multiple
  electric charges,  in 13~\tev $pp$ collisions. The solid lines are leading-order cross-section calculations.} 
\end{figure*}

 \begin{table*}[!htb]
\caption{\label{tab:masslimits} 95\% CL mass limits found in a Drell-Yan production model for spin-0, \mbox{spin-\half} and spin-1 dyon pair direct production in LHC $pp$ collisions, assuming $\beta$-independent couplings. 
} 
\vspace{0.2cm}
\begin{tabular}{c|c|cccccccccccccccccc}
Magnetic & Spin  & \multicolumn{18}{c}{Electric charge}  \\
 charge   &           & 0         & 1         & 2        & 3        & 4        & 5        & 6        & 10      & 15      & 20     & 25      & 50          & 75    & 100    & 125   & 150 & 175 & 200 \\ 
\colrule
& & \multicolumn{18}{c}{95\% CL mass limits [GeV]  } \\
\colrule
1$g_D$  &  0        & 870  & 750   & 750   & 750   & 750   & 750   & 750   & 760  & 780   & 790 &  810  & 920     & 970  & 1010 & 1010 & 1000 & 950 & 840 \\
2$g_D$  &   0       & 1240& 1110  & 1110 & 1110 & 1110  & 1110  & 1110  &1120 & 1120 & 1110 & 1110 & 1100 & 1070 & 1050 & 1020 & 950 & 870 & 770  \\
3$g_D$  &   0       & 1300 &1170  & 1170 & 1170 &1170  & 1170 & 1160  &1170 & 1170 & 1160 &1160  & 1130  & 1070 & 1020 &  940 &  850 &     -     &   -    \\
4$g_D$  &   0       & 1200 &1060  & 1060 & 1060 &1060 & 1060 & 1060  & 1060 &1060 &1040 &1030   & 970  & 900* & -         &   -    &   -       &     -     & -  \\
\colrule
 1$g_D$  &  \half  &  1410 & 1270 &1270 & 1270 & 1270 & 1270 & 1260 &1280 &1300  &1310 & 1330   & 1400  & 1450 &  1460  & 1420 & 1370 & 1310 &  -  \\
 2$g_D$  & \half    &  1810 & 1660 &1660 & 1660 & 1660 & 1660 & 1650 & 1650 &1650 & 1640 & 1640   & 1600 & 1550 &  1500 & 1420 & 1350 &  -        & - \\
 3$g_D$  &  \half   &  1840 & 1700 &1700 & 1690 & 1690 & 1680 & 1680 & 1670 & 1670 &1650 & 1640   & 1580 & 1500 &  1420 & 1330 & 1280 & -         &  - \\
4$g_D$  &  \half   &  1680 & 1560 &1550 & 1550 & 1540 & 1540 & 1540 & 1530 & 1520& 1510 & 1480   & 1420  & 1340 & 1280 &  1210 &    -      &  -        &  -\\
5$g_D$   &\half    &  1460 & 1300 &1300 & 1300 & 1300 & 1300 & 1310 & 1300 &1290 & 1290 & 1290  &  1220  & -          &      -     &            &   -       &           & \\
\colrule

1$g_D$  & 1        & 1460 & 1340 & 1340 & 1350 & 1350 & 1350 & 1340 & 1360 & 1390 & 1420 & 1450 & 1550   & 1620  &  1670 & 1670 &1650  & 1610 & - \\
2$g_D$  &  1       & 1930 & 1790 & 1790 & 1790 & 1790 & 1790 & 1790 & 1800 & 1800 & 1790 & 1790 & 1780   & 1770 &  1740  & 1710 & 1640 & 1590 &1520  \\
3$g_D$  &  1       & 2040 & 1910 & 1910 & 1910 & 1900 & 1900 & 1900 & 1900 & 1890 & 1890 & 1890 & 1840   & 1790 &  1730 &  1670 & 1600 &   -  &  - \\
4$g_D$  &  1       & 1990 & 1860 & 1850 & 1850 & 1840 & 1840 & 1840 & 1840 & 1830 & 1820 & 1810 & 1750  &  1690 &  1620  & 1570 &     -    &   -   &   -  \\
5$g_D$   & 1       & 1820 & 1690 & 1680 & 1680 & 1670 & 1660 & 1670 & 1650 & 1660 & 1640 & 1640 & 1570  &  1500 & 1480  &  -      &  -        &     -   &   -     \\

\end{tabular} 
\end{table*}

We calculate cross-section upper limits at 95\% CL using as benchmark a DY model for dyon and magnetic monopole production, assuming
 a $\beta$-independent coupling, for  three spin hypotheses (0, \half, 1), magnetic charge up to 5$\gd$ and in the dyon case, electric charge up to 200$e$. These values mark the limit of the sensitivity of this search due to the absorption of higher charges in the material comprising LHCB's VELO detector that lies between the IP and the MMT detector.
 
 An example of the limit curves obtained for spin-1/2 dyons with charge 1$\gd$ are  shown   in Figure~\ref{fig:limits1}. The corresponding limits for other dyons with  spin-0, spin-1/2 and spin-1  and magnetic charges ranging up to and including $5\gd$ are  given in the supplemental material \cite{supplemental}. They are extracted on the basis of the acceptance estimates and their uncertainties; the delivered integrated luminosity 6.46~$\ifb$,  measured a with a precision of 4\% \cite{Aaij:2018}, corresponding to the full 2015-2018 exposure to 13~\tev $pp$ collisions 
 and the non-observation of magnetic charge inside the trapping detector samples.

 Using cross-sections, computed at leading order,  mass limits are obtained and reported in Table~\ref{tab:masslimits}.  It is important to note that these 
 DY cross-sections  are computed using perturbative field theory. However, the monopole-photon coupling is too large for such  an approach. Thus, the mass limits given are only indicative.

 Comparing the dyon mass limits with the corresponding  monopole mass limits \cite{Acharya:2019}  obtained from the same dataset using an analogous we find, not surprisingly, that for the smallest electric charge of the dyon (1$e$) the limits  obtained are comparable or better than those obtained  in the  monopole search, as indicated in Table~\ref{tab:masslimits}. 

In summary, we considered the direct  production of dyon-antidyon pairs via the  DY mechanism for the first time at an accelerator. The aluminium elements of the MoEDAL trapping detector exposed to 13~\tev LHC collisions during Run-2 were scanned using a SQUID-based magnetometer to search for the presence of trapped magnetic charge belonging to dyons. No candidates survived our scanning procedure and cross-section upper limits as low as 30~fb were set. As mentioned above, the trapping condition requires the dyon to be negatively electrically  charged. Mass limits in the range 750 - 1910~ GeV were set using a benchmark DY production model,  for dyons with magnetic charge up to 5$\gd$, for electric charge from 1$e$ to 200$e$, and, for spins~0, \half and~1. The  corresponding mass limits  for magnetic monopoles are in the range 870 - 2040~GeV for magnetic charges in the same range.   

We note  that many previous searches for highly ionizing particles would  in principle also have sensitivity to dyons. However, no explicit search for dyons has ever been performed to date. We suggest that dyons be added to the list of highly-ionizing particles for which dedicated searches are conducted at the LHC and at future colliders.

 \vspace*{-3mm}
\section{Acknowledgments}
We thank CERN for the  LHC's successful  Run-2 operation, as well as the support staff from our institutions without whom MoEDAL could not be operated. We acknowledge the invaluable assistance of  particular members of the LHCb Collaboration: G. Wilkinson, R. Lindner, E.  Thomas and G. Corti. Computing support was provided by the GridPP Collaboration, in particular by the Queen Mary University of London and Liverpool grid sites. This work was supported by grant PP00P2\_150583 of the Swiss NSF; by the UK Science and Technology Facilities Council, via the grants ST/L000326/1, ST/L00044X/1, ST/N00101X/1, ST/P000258/1 and ST/T000759/1; by the Generalitat Valenciana via a special grant for MoEDAL and via the projects PROMETEO-II/2017/033 and PROMETEO/2019/087; by MCIU / AEI / FEDER, UE via the grants FPA2016-77177-C2-1-P, FPA2017-85985-P, FPA2017-84543-P and PGC2018-094856-B-I00; by the Physics Department of King's College London; by  NSERC via a project grant; by the V-P Research of the University of Alberta (UofA); by the Provost of the UofA); by UEFISCDI (Romania); by the INFN (Italy);by the Estonian Research Council via a Mobilitas Plus grant MOBTT5; and by a National Science Foundation grant (US) to the University of Alabama MoEDAL group.
 \vspace*{-3mm}

\bibliographystyle{revtex4-2}

\end{document}
